\documentclass[]{article}
\usepackage{lmodern}
\usepackage{amssymb,amsmath}
\usepackage{ifxetex,ifluatex}
\usepackage{fixltx2e} % provides \textsubscript
\ifnum 0\ifxetex 1\fi\ifluatex 1\fi=0 % if pdftex
  \usepackage[T1]{fontenc}
  \usepackage[utf8]{inputenc}
\else % if luatex or xelatex
  \ifxetex
    \usepackage{mathspec}
  \else
    \usepackage{fontspec}
  \fi
  \defaultfontfeatures{Ligatures=TeX,Scale=MatchLowercase}
\fi
\IfFileExists{upquote.sty}{\usepackage{upquote}}{}
\IfFileExists{microtype.sty}{%
\usepackage{microtype}
\UseMicrotypeSet[protrusion]{basicmath} % disable protrusion for tt fonts
}{}
\usepackage{hyperref}
\hypersetup{unicode=true,
            pdfborder={0 0 0},
            breaklinks=true}
\ifx\paragraph\undefined\else
\let\oldparagraph\paragraph
\renewcommand{\paragraph}[1]{\oldparagraph{#1}\mbox{}}
\fi
\ifx\subparagraph\undefined\else
\let\oldsubparagraph\subparagraph
\renewcommand{\subparagraph}[1]{\oldsubparagraph{#1}\mbox{}}
\fi

\date{6 June 2012}

\title{Deploying Static Analysis}\label{deploying-static-analysis}
\author{
Flash Sheridan \\ Bell Labs/Orin USA \\ \textit{\url{http://pobox.com/~flash}}
\\ \\
This is the original unabridged version (with footnotes) of the \\
\textit{Dr Dobb’s Journal} August 2012 cover story.
}  

\begin{document}

\maketitle

\section{}\label{section-1}

\section{}\label{section-2}

\section{}\label{section-3}

\begin{abstract}\label{abstract}\normalsize 
Static source code analysis is a powerful tool for finding and fixing
bugs when deployed properly; it is, however, all too easy to deploy it
in a way that looks good superficially, but which misses important
defects, shows many false positives, and brings the tool into disrepute.
This article is a guide to the process of deploying a static analysis
tool in a large organization while avoiding the worst organizational and
technical pitfalls. My main point is the importance of concentrating on
the main goal of getting bugs fixed, against all the competing lesser
goals which will arise during the process.
\end{abstract}

\subsection{Introduction}\label{introduction}

Static source code analysis for bug finding (``static analysis'' for
short) is the process of detecting bugs via an automated tool which
analyzes source code without executing it. The idea goes back at least
to Lint, invented at Bell Labs in the 1970's, but underwent a revolution
in effectiveness and usability in the last decade, so much so that
``Lint'' is now sometimes used as a term of abuse {[}Engler et al
2010{]}, p. 75. The initial focus of static analysis tools was on the C
and C++ programming languages, as is my own background; such tools are
particularly necessary given C/C++'s notorious flexibility and
susceptibility to low-level bugs. More recently, tools have flourished
for Java and/or web applications; these are needed because of the
prevalence of easily exploitable network vulnerabilities.

Leading commercial static analysis tools with which I am familiar
include Coverity, Fortify (now owned by Hewlett-Packard), and Klocwork.
Klocwork and Coverity both initially focused on C/C++, though they came
from opposite origins: Klocwork from the telephone equipment company
Nortel, and Coverity from Stanford University. Fortify's initial focus
was on security for web applications in languages such as Java and PHP.
All three companies are now encroaching on each other's territories, but
it remains to be seen how well they will do outside of their core
competencies.

An excellent, free, but limited, academic static Java byte code analysis
tool is FindBugs. Its lack of an integrated database for defect
suppression makes its large-scale use difficult in sizeable
organizations, but its use by individual developers within the Eclipse
development environment can be extremely valuable. Similarly, recent
versions of Apple's Xcode and Microsoft's Visual Studio development
environments contain integrated static analysis tools for C/C++. These
are useful for finding relatively shallow bugs while an individual
developer is writing code; their short feedback loop avoids the
difficulties discussed in this paper, with broader deployment of tools
which perform deeper analysis. A longer list of tools is provided by the
Gartner Magic Quadrant Analysis {[}Gartner 2009{]}, albeit with a strong
bias towards security and adherence to Gartner's strategy
recommendations.

There is no general-purpose introductory textbook on the subject; the
best general introduction is the previously cited short article by
Dawson Engler, the inventor of Coverity. Two of the leaders at Fortify
have written an introductory textbook, but it focuses primarily on their
tool and on security, and skimps on key static analysis concepts
{[}Chess \& West 2007{]}. There is a rigorous academic textbook on the
more general sense of static analysis, for purposes not restricted to
bug finding, but it preceded the revolution in static analysis for bug
finding {[}Nielson et al. 1999{]}. At least two prominent academics are
rumored to be writing books on static analysis, but neither has anything
to say publicly yet.

\subsubsection{ Purpose}\label{purpose}

The first question to ask before deciding to do static analysis in an
organization is not what tool to buy, nor even whether you should do
static analysis at all. It's ``why?''

If your purpose is genuinely to help find bugs and get them fixed, then
your organizational and political approach must be different from the
more usual, albeit unadmitted, case: producing metrics and procedures
that will make management look good. (Fixing bugs should actually be
your \emph{second} goal: an even higher goal is preventing bugs in the
first place, by making your developers learn from their mistakes. This
also contraindicates outsourcing evaluation and fixing of defects,
tempting though that may be.)

\subsection{ Political Issues to Settle in
Advance}\label{political-issues-to-settle-in-advance}

Get buy-in from, and about, your testing/quality assurance department:
that they support the project, and will have authority over
quality-related issues, even if they inconvenience the other
stakeholders. Quality has a much smaller constituency than the schedule
or the smooth running of internal procedures, but it must be the final
arbiter for crucial quality-related decisions {[}Spolsky 2004{]}, ch.
22, pp. 171--8.

\subsubsection{Management of the Tool}\label{management-of-the-tool}

Give some thought to what part of the organization, if any, should be in
charge of running the tool once it's set up. If your organization has a
tools team, they may seem the obvious owners, but this does need careful
consideration. Static analysis for bug finding is probably not your
organization's core competency, and you will need to worry about the
Iron Law of Bureaucracy: your tools team's institutional interest will
be in the smooth running of the tool, not in the messy changes necessary
for finding bugs. Even if you're reluctant to outsource the rest of the
process, administration and configuration may be more flexible if done
by external players rather than an internal team with its own interests,
habits, and procedures. It may also be more productive to hire an
expensive consultant for a few hours, rather than a lesser-paid internal
resource full-time; an external resource may be more flexible and less
prone to establishing an entrenched bureaucracy.

\subsubsection{Engineering Management}\label{engineering-management}

The conventional wisdom is that getting most developers to use a static
analysis tool requires a high-ranking management champion (mentioned in
{[}Engler et al. 2011{]}, but a much older idea than that), to preach
its benefits, ensure that the tool is used, and keep the focus on
finding bugs and getting them fixed. The flip side to this is that any
attempt to herd cats, i.e., to get programmers to adhere to best
practices, will cause a backlash. So your tool must withstand scrutiny
from developers looking for excuses to stop using it.

Get buy-in from engineering that they will make time in the schedule to
review and fix bugs found, even if they are disinclined to do so, which
they will be once they see the first false positive. (Or even the first
false false positive; more on this below.) Ensure that it's not the
least effective engineers whose time is allotted for reviewing and
fixing static analysis bugs---see ``Smart Programmers Add More Value,''
below.~~You'll also need agreement from the security team that salesmen
and sales engineers will get access to your real source code; more on
that below.

\subsubsection{Smart Programmers Add More Value---and Subtract
Less}\label{smart-programmers-add-more-valueand-subtract-less}

Handling static analysis defects is not something to economize on.
Writing code is hard, finding bugs in professional code \emph{should} be
hard, and evaluating possible mistakes in alleged bugs is even harder.
Learning to evaluate static analysis defects, even in a developer's own
code, requires training and supervision. It is necessary to tread
delicately around the polite pretence that the code owner is an
infallible authority on the behavior of that code. Misunderstandings
about the actual behavior of unsigned integers and assertions are, for
instance, regrettably common in my experience. See {[}Engler et al
2010{]} p. 73 for further examples.

It can be tempting to delegate the triaging (i.e., initial screening) of
static analysis defects, at least initially (at least in theory), to
someone other than the code owner. This is much more difficult and
hazardous than at first appears. The abstract (and odd)
language-lawyerly mindset required is two levels of abstraction higher
than that of a practical programmer, and socially useful attitudes that
promote teamwork can work against the correct understanding of defects.
Negative capability, realizing when only the code's owner can decide
some issues, is also rare but necessary. My rule of thumb is that a
small proportion of developers grasp static analysis quite quickly, but
that there is about a fifty percent chance that the rest will reach an
accuracy of fifty percent after about a month of supervised triaging.

It's quite easy for unsupervised triagers to subtract much of the value
you could get from static analysis. The conventional wisdom strikes with
a vengeance: a smart programmer adds more value than several cheap
programmers, and subtracts vastly less value {[}McConnell 1996{]}, p.
12. If an unqualified triager triages half of the genuine defects away
as false false positives, and marks half the false positives as genuine,
then the engineer responsible for fixing them won't see many of the
genuine defects, and may use the falsely-marked false positives as an
excuse to speed-mark the rest as false positives too.

Even more importantly, the most valuable benefit from static analysis,
greater even than fixing bugs, is preventing future bugs, by educating
the developer about his or her mistakes. If someone else is looking at
these bugs, the developer never sees the mistakes and cannot learn from
them.

\subsubsection{Not Our Problem}\label{not-our-problem}

If your software included third-party code, it's an unpleasant political
reality that you may include it in your product without fixing it.
Whether this is a good idea, for your customers and/or your
organization, is outside the scope of this article; but if you're not
going to fix it, then you shouldn't spend too much time investigating
static analysis defects in it. On the other hand, you \emph{should} nag
your suppliers to use static analysis to find and fix their bugs.

\subsection{Deciding Which Tool to Buy: Do a Real
Test}\label{deciding-which-tool-to-buy-do-a-real-test}

The good news about evaluating expensive static analysis tools is that
most of them have free demos, and you don't have to give the bugs back
for tools you don't buy. This suggests that the vendors are confident
that they'll find many serious bugs in your code; that they haven't gone
out of business suggests that they're right.

\subsubsection{Salesman's Proof of
Concept}\label{salesmans-proof-of-concept}

A demo will probably entail giving a sales engineer access to your
source code and build system, either on-site or with remote access. Be
prepared for internal resistance; this level of trust will be troubling
to your security experts, but there's no serious alternative:
experimentation with your build system may be necessary.

Avoid the temptation to cut corners and scan only what's convenient;
it's very easy to waste your time completely unless you do a real test
on your real code, and detect serious bugs that need to be fixed. The
upside is that if you do this right and the tool is worth buying, it
will do a good job of selling itself, by exhibiting bugs that convince
even high-ranking sceptics that the tool is worth the price.

You may need to accept temporarily an inconvenient ratio of false
positives (provided they can be cleaned later up by configuration
tuning), and unpolished integration into your build system. Polishing
the system beyond the point where you decide to buy it is not the sales
engineer's job; that should be done after the sale, by the vendor's
support engineers. This will require a fair amount of judgment, to tell
that an unpolished proof of concept can lead to a production-ready
system.

\subsubsection{Deciding Among Demos}\label{deciding-among-demos}

Deciding which tool has demo'd best should be straightforward: Which one
has found bugs in your real code that will convince management to spend
serious money and resources? Ease of use, including a low false-positive
ratio, is an important secondary consideration. Resist the temptation to
compromise by buying more than one tool at first. One tool which is
properly deployed will find more (or better) bugs (or fewer false
positives) than two tools which are spread too thin and become
shelf-ware.

In the long run, once you've found and fixed all the bugs your first
tool can find, by all means consider buying and installing another
static analysis tool: The conventional wisdom is that static analysis
tools have depressingly little overlap. But this may be in the
\emph{extremely} long run; it's also conventional wisdom that a good
static analysis tool will find more serious bugs than you're willing to
allot resources to fix (see below).

\subsubsection{Real Installation}\label{real-installation}

Once you've actually bought a static analysis tool, be much fussier
about the real setup than you were about the demo. Once the installation
engineer(s) consider themselves finished, you won't have much leverage
for major changes, and will be dependent on remote support personnel,
who are lower on the pecking order.

Discuss how you're going to rank defects with the installation engineers
before you leave, but resign yourself to handling this later yourself
(see below). Ask around about the support personnel, and try to
negotiate to have your requests handled by the better ones. Do read the
documentation; it's unlikely to be complete, but the user interfaces for
these tools are not well designed, with essential hidden functionality,
and sometimes regress in both features and quality. It may be worth
running an older but better-tested version of the tool.

\subsection{ Politics and Procedure: Preparing for an Embarrassment of
Riches}\label{politics-and-procedure-preparing-for-an-embarrassment-of-riches}

The Embarrassment of Riches problem means that a modern commercial
static analysis tool generally finds more bugs than the user has
resources, or at least willingness, to fix {[}Sheridan 2010b{]}.
Political resistance to static analysis bugs is sometimes warranted (see
{[}Pugh 2009{]}), sometimes mere laziness, but sometimes deeper and
cultural: Avoiding the kinds of bugs that static analysis finds is
largely a matter of discipline, which is unpopular (sometimes
justifiably) among most programmers. Fixing these bugs, and verifying
that your organization has done so, will require adaptability and
judgment. Attempts to design simple rules and metrics for this are, in
my opinion, at best premature, and perhaps impossible.

\subsubsection{Non-Goals and Metrics}\label{non-goals-and-metrics}

\paragraph{Two Fallacies}\label{two-fallacies}

Beware of two fallacies. The first is one of the standard risks of
measurement: over-optimization of single factor measurement {[}McConnell
1996{]} §26.2 p. 476, {[}Spolsky 2004{]}, §28, p. 211; the factor you
measure may be all that gets optimized, displacing efforts towards your
genuine goal. The second is what I call the Management Metrics Fallacy:

\emph{If it can't be measured, it can't be managed.}

\emph{$\therefore$ What's easy to measure is all that's important.}

Stable metrics are the enemy of quality: If your tool must produce
numbers that won't oscillate or upset people, then it can't change
rapidly in order to catch real bugs, or to stop reporting fake ones.
This will bring static analysis into disrepute within your organization,
and give engineers an excellent excuse for ignoring (or de-prioritizing)
static analysis defects.

Conversely, it's easy to track a number that makes management look good
but doesn't get important bugs fixed, e.g., the number of projects
analyzed, or the number of unexamined defects. And even though getting
bugs fixed is the goal, simply tracking that number may still be
misleading (I have been guilty of this myself). If you put too much
emphasis on the number of defects fixed, the developers being measured
may spend more time than is warranted on unimportant (or even obsolete)
areas of the code, where static analysis defects may be more common.
Some code should be excluded from analysis and ignored, rather than
fixed. And, sadly, some shipping code will be excluded from being fixed
(see ``Not Our Problem,'' above).

\paragraph{What Counts}\label{what-counts}

What counts is a hypothetical, and hence impossible to measure with
certainty: How many serious bugs did you prevent from reaching
customers? This is related to a secondary consideration: How many bugs
did you prevent from reaching manual testing, when bugs start to get
expensive?

\subparagraph{Return on Investment}\label{return-on-investment}

Demonstrating systematically that a static analysis tool has been worth
its cost, both in money and (more importantly) engineering time and
effort, seems to be an unsolved problem. Metrics abound, but they're
generally subject to sceptical objections, e.g., that the bugs were in
unimportant code, or couldn't have been very important if they weren't
noticed during testing. Anecdotal evidence is generally more solid,
e.g., particular bugs whose impact is obvious, though it can be
surprisingly hard to close the loop on this. I've only managed it
completely once, with a defect which was reported by a static analysis
tool but ignored, and later found and escalated in user-level testing,
and then debugged. (I discovered the connection through watching source
code check-ins, and changes in the status of defects found by the tool.)
There are of course occasional retroactive detections of high-publicity
bugs, but my gut feel is that these are the exceptions rather than the
rule, and that the bulk of the benefit is in finding bugs which would be
caught later but more expensively. The picture may be radically
different for security-related issues, but data on that is even scarcer.

\paragraph{What Doesn't}\label{what-doesnt}

Failing to distinguish bugs which have been fixed, from bugs that merely
were in functions or files which have been removed, is a common
shortcoming in static analysis tools, but is a substantial obstacle to
measuring how much good a static analysis tool is doing.

Smooth running of the tool is another a non-goal and a potential
obstacle: If someone finds a problem with the configuration, it needs to
be changed quickly, before it wastes more time and misses more genuine
bugs, even if this requires broad-reaching configuration changes.
Finding bugs is inherently messier, and requires much more flexibility,
than writing code or maintaining a tool; this can be a source of acute
cultural conflict. A change review process which makes sense for
production code can stymie bug finding, waste engineering time on false
positives, and bring the tool into disrepute with those actually using
it.

\subsubsection{Be Prepared for False
Negatives}\label{be-prepared-for-false-negatives}

No static analysis tool will find all bugs in any significant code base;
part of the revolution in static analysis was giving up on even limited
attempts to do so, in favor of heuristics to find actionable bugs.
Definite numbers are hard to come by, but Coverity's Analysis Architect
estimates that it probably finds less than 20\% of bugs
present.\footnote{Roger Scott, Coverity Analysis Architect, in
  LinkedIn Static Code Analysis group, 6 December 2011,
  \textless{}\url{http://www.linkedin.com/groupItem?view=\&gid=1973349\&type=member\&item=81776104\&qid=2ca8f00f-2aea-422b-9b45-dfb315e96cae}\textgreater{},
  registration required.} It seems unlikely that any current tool can
do much better except at the cost of an unwieldy number of false
positives. (Vendor claims of completeness are generally so restricted as
to be impractical for normal development.) This is less of a
disadvantage than it seems, since the Embarrassment of Riches problem
means that a modern commercial static analysis tool generally finds more
bugs than the user has time to fix. Thus prioritization will be crucial.

\subsubsection{Living with False
Positives}\label{living-with-false-positives}

Conversely, no significant static analysis tool is immune to false
positives; the number of these tends also to be, very roughly, 20\%.
This number, however, is usually eclipsed by the number of technically
correct defects which are of little interest for other reasons, such as
being in a part of the code that no-one cares about, or being (at least
in the opinion of the code owner) unlikely to occur in the field. (You
must also be prepared for a high false false positive ratio, though this
varies widely among developers.) Even with a high false positive (or
don't care) ratio, static analysis is still vastly more efficient than
other forms of bug finding, since it takes little time for a code owner
to dismiss an irrelevant defect. (Triaging false positives can be more
time-consuming for people not familiar with the code, however.) It is
important to manage expectations, nonetheless, so that someone tasked
with examining static analysis defects is not discouraged by false
positives from persevering in dealing with real bugs.

\subsubsection{Ranking and
Prioritization}\label{ranking-and-prioritization}

Most static analysis tools present defects in essentially random order
(e.g., alphabetical by checker name, or by file), which is unwise: If
the first defect in a given engineer's queue is unimpressive, you may
have lost him {[}Engler \& Kremenek 2003{]}. This is particularly
disappointing since there are techniques in the academic literature for
ranking defects by reliability, relevance, and estimated importance
{[}Engler \& Kremenek 2003{]}, {[}Engler \& Kremenek 2004{]}.

One common but insufficient facility is ranking defects by the checker
which finds them: An unreliable defect with little impact is less
important than a definite bug with bad consequences, regardless of which
checkers found them. I have no general solution to offer users of
existing tools, beyond the advice below on experimenting, measuring, and
adapting. Don't dismiss even unambitious checkers too hastily. What I
call \emph{Engler's Third Law} is that no bug is too foolish to check
for {[}Engler et al 2010{]} p. 75, and this is depressingly well
confirmed by experience, at least for C/C++. (Egregious bugs may be
harder to find in Java.) In under-development proprietary code, the
simplest checks for the most painfully obvious bugs are often the most
effective, e.g., use immediately after free, and a culture of not
null-checking memory allocation, nor freeing resources on error paths.
More sophisticated checkers, e.g., for concurrency errors, can find
impressive bugs, but may also have a high don't care (or don't
understand) rate.

\subsection{ Configuration: on Your
Own}\label{configuration-on-your-own}

\subsubsection{Improper configuration and build
errors}\label{improper-configuration-and-build-errors}

Improper configuration, in particular errors in locating and
meta-compiling your source, can silently ruin your analysis while
leaving the illusion of doing useful work. Build and parsing errors can
make most defects false positives, or miss most of the genuine defects
the tool is capable of finding. The upside is that a few simple (though
perhaps hard-to-find) fixes to the build configuration can convert a
worse-than-useless analysis, consisting of mostly false positives, into
something worth getting your engineers to look at and fix bugs with. But
you must set up the procedure so that those configuration fixes are made
promptly: Letting a bogus build count as a success, or setting up a slow
bureaucracy to approve changes, will bring the tool into disrepute, and
give engineers an excuse to ignore even legitimate bugs found by the
tool.

If your tool has a minimum threshold for reporting success of an
analysis, set it very near to the most fussy. If not, get agreement that
a certain ratio of errors per line will invalidate an analysis run.
(Better still, make the build script count errors and stop if the
threshold is exceeded.) Otherwise your tools team can report success no
matter how bad the analysis is.

Diagnosing build errors often hinges on the values of compile-time
macros; in a complex build system, such definitions may be nested in
confusing ways. It is not necessarily obvious (at least to me) how to
tell the value of such a macro. One trick, for some compilers, is to
generate a compile time error for a bogus \#include file whose name
includes both the name of macro and its value, as in the following. (The
output is messy but includes the needed information.)

\#define COMPILE\_TIME\_PRINT\_AND\_STOP(x) \textless{}
"COMPILE\_TIME\_PRINT\_AND\_STOP:" \#x\#\#: x \textgreater{}

\#define \emph{\textbf{\_\_dest\_os Alien/OS}}

\#include COMPILE\_TIME\_PRINT\_AND\_STOP(\emph{\textbf{\_\_dest\_os}})

\subsubsection{Better Configuration: Enable the Good Checkers and
Disable the
Cruft}\label{better-configuration-enable-the-good-checkers-and-disable-the-cruft}

The flip side of fixing a broken build is improving a good one.
Investigate and experiment with your settings on \emph{your} code. The
vendor's defaults are likely to be a one-size-fits-all configuration,
designed to make the tool look good and minimize problems for their
support staff. The vendor may have disabled some checkers because of a
high rate of false positives in rare circumstances. Conversely, some
checkers may be unimpressive on your code. Some have statistical
thresholds, which need to be adjusted for your group's coding culture.

\subsection{Handling Static Analysis Defects
}\label{handling-static-analysis-defects}

\subsubsection{Bugs Found and Fixed}\label{bugs-found-and-fixed}

Identifying bugs, and their fixes, via static analysis is the easy part;
the tool goes right to the heart of a defect, and highlights the problem
in clear, bright colors. Difficulties usually arise after the fix is
identified, either because it is misunderstood (which I discuss below),
or because there is resistance to making the change, which is largely
outside the scope of this paper. (Sometimes this resistance is even
justified {[}Pugh 2009{]}.) Given that a tool is good enough to find
real bugs, what determines whether the bugs get \emph{fixed} is the
sociology and politics.

\subsubsection{Bugs Missed or Ignored}\label{bugs-missed-or-ignored}

The conventional wisdom is that static analysis tools have depressingly
little overlap; but missing bugs is not the end of the world. As long as
the pipeline is full of genuine and significant bugs, a good static
analysis tool will do a lot of good for the code and the customers.

Conversely, some genuine bugs found by the tool will be ignored, either
as false false positives, annoyances swept under the rug, or simply not
gotten around to in time, due to the Embarrassment of Riches problem.

\subsubsection{Misclassified Bugs: False False Positives and Other
Mistakes}\label{misclassified-bugs-false-false-positives-and-other-mistakes}

Keep a close eye on the ratio of false positives; anything much above
20\% indicates a configuration error, user errors, or (more rarely) a
bad checker which should be disabled.

Novice static analysis users need to accept that, when they disagree
with the tool, they will usually (but not always) be wrong. This is not
to say that any tool is perfect; once a user understands the kinds of
mistakes it makes (and doesn't make), human supervision can begin to add
value. So do not make the opposite mistake and fix all defects blindly;
unnecessary code changes carry their own risks, and I have seen mistakes
(detected by later static analysis) in code required to silence earlier
static analysis warnings. In particular, a strict policy of requiring
immediate fixes for newly checked-in static analysis defects might have
unintended consequences. A programmer who is in a hurry to get home
after a check-in is unlikely to be in the best state of mind for
analyzing and fixing static analysis defects.

A grayer area is defects where the tool is technically correct, but the
developer believes that the code path is not worth worrying about. This
is a judgment call, and hence hard for anyone but the code owner to
second-guess; this also makes it impractical to judge statistically.

\subsubsection{Practicalities}\label{practicalities}

Most of the major tools present their end-user interface for defects in
a web browser. This makes the user interface of your particular browser
of considerable importance, and requires a delicate balance between
habitual usability and specific features needed for examining static
analysis defects.

\paragraph{Symbol Highlighting}\label{symbol-highlighting}

Firefox's Highlight All feature is invaluable for quickly highlighting
all occurrences of the currently selected text, e.g., a variable or
function name.If you will be examining a number of defects, it's worth
practicing the key sequence until it's instinctive. (On Macintosh
Firefox, one of the keyboard equivalents is missing, so a keyboard macro
program is necessary.)

Most browser-based tools have similar functionality that's symbol-based,
which has advantages, disadvantages, and limitations. Some instances of
the symbol can be missed, but on the other hand, a symbol which is a
superstring of the selected symbol will be properly ignored by
symbol-based search, but improperly highlighted by a browser's
string-based search.

Quick highlighting can reduce some types of defects from a cognitive
problem to simple pattern recognition. For instance, a buffer overrun
defect with a parameter array index can often be recognized merely by
highlighting the index on the line where the tool reports the defect.
The pattern of color between there and the function head makes it
obvious where there are checks, if any, on the parameter.

\paragraph{Incompatibilities and
Difficulties}\label{incompatibilities-and-difficulties}

Some tools cause difficulties with some browsers. Coverity 5.x's
heavy-duty web pages, for instance, do not work properly with some
versions of Firefox, and will not open separate pages for function
definitions; this is not a problem with Coverity 4.x. (My workaround for
the latter limitation is to take quick notes via text drag and drop, but
this requires support from both the browser and your text editor.)
Fortify's web interface uses Adobe Flash, which does not support
standard text selection behavior, so I recommend avoiding it in favor of
their standalone application (which has text interface difficulties of
its own.)

\paragraph{Defect Tracking}\label{defect-tracking}

Most commercial tools come with a defect-tracking system, which is
essential for suppressing false positives. This is essentially a bug
database, but without many essential features, such as status tracking.
(The first thing a quality assurance professional checks in the morning
is changed bugs in the database, for instance to see what developers and
managers have been marking as unimportant.) None of the bundled
databases known to me matches the features of professional bug-tracking
databases; I have sometimes been reduced to archiving bug lists as CSV
files, and diffing them as time goes by.

Sometimes it is useful to link an organization's real bug-tracking
database with a static analysis defect database. I don't advise
importing defects automatically; in my opinion, a real bug database
should be reserved for issues which a human has judged to be genuine
bugs. One strategy I have found useful is to file one bug report per
component into your existing bug database, with a link in each bug
report to a live query with suspected defects in the component in the
\emph{current} build, with separate counts for high- and low-priority
checkers.

\subsubsection{Keep Experimenting, Measuring, Adapting, and
Educating}\label{keep-experimenting-measuring-adapting-and-educating}

As I have mentioned above, metrics---applied blindly---make a very bad
master; but applied wisely, they can be a helpful servant. Track the
numbers for important (and unimportant) defects found, and the checkers
which found them. Even valid checkers may not find the sort of defects
which your organization is, realistically, going to fix. Track the
defects which actually get fixed, preferably by keeping an eye on source
code management check-ins. (Pretend not to see check-in comments which
minimize the importance of the defect fixed; developers are often
reluctant to admit that an automated tool found significant bugs in
their code. These imperfect developers are your best customers, not your
enemy.) Given finite resources, it is crucial to prioritize defects;
sadly, this sometimes means ignoring some components, or disabling some
checkers.

Also keep an eye on static analysis defects which are \emph{not} fixed.
Sometimes the problem will be with the tool, and some checkers are not
worth the resources they consume. More often the problem will be with
the reaction to the tool, and sometimes education is necessary. That is
perhaps the most valuable result of static analysis: Making developers
think about their code, and learn about what it \emph{actually} does, is
in the long run even more important than fixing the current version of
their code.

\subsection{ References}\label{references}

\begin{itemize}
\item
  Brian \emph{Chess} and Jacob West, \emph{Secure Programming with
  Static Analysis,} Addison-Wesley 2007.
\item
  Ted Kremenek and Dawson \emph{Engler}, ``Z-ranking: using statistical
  analysis to counter the impact of static analysis approximations,''
  \emph{Proceedings of the 10th Annual International Static Analysis
  Symposium,} Springer-Verlag 2003, pp. 295-315.
\item
  Ted Kremenek, Ken Ashcraft, Junfeng Yang, and Dawson \emph{Engler},
  ``Correlation exploitation in error ranking,'' \emph{SIGSOFT Software
  Engineering Notes 29,} 6 (Nov. 2004), pp. 83-93.
\item
  Al Bessey, Ken Block, Ben Chelf, Andy Chou, Bryan Fulton, Seth Hallem,
  Charles Henri-Gros, Asya Kamsky, Scott McPeak, and Dawson
  \emph{Engler}, ``A few billion lines of code later: using static
  analysis to find bugs in the real world,'' \emph{Communications of the
  ACM,} volume 53, number 2 (2010), pp. 66-75.
\item
  J. Feiman and N. MacDonald, ``Magic Quadrant for Static Application
  Security Testing,'' \emph{Gartner} RAS Core Research Note G00208743,
  13 December 2010.
\item
  Steve \emph{McConnell}, \emph{Rapid Development,} Microsoft Press
  1996.
\item
  Flemming \emph{Nielson}, Hanne R. Nielson, Chris Hankin,
  \emph{Principles of Program Analysis,} Springer-Verlag 1999.
\item
  Bill \emph{Pugh}, ``Cost of static analysis for defect detection,''
  unpublished talk at Stanford University Computer Science Department,
  20 April 2009.
\item
  Flash \emph{Sheridan}, ``Handling an embarrassment of riches,'' Code
  Integrity Solutions blog posting, 29 January 2010,
  http://codeintegrity.blogspot.com/2010/01/handling-embarrassment-of-riches.html.
\item
  Flash \emph{Sheridan}, ``Static analysis deployment pitfalls,''
  \emph{Supplemental Proceedings of the 21st IEEE International
  Symposium on Software Reliability Engineering,} November 2010.
\item
  Joel \emph{Spolsky}, \emph{Joel on Software,} Apress 2004.
\end{itemize}

\emph{\small{©2011--2 Flash (K.J.) Sheridan.  Converted to LaTex 30 May 2016; heading and formatting tweaks 11 July 2021 and 30 January 2022.}}

\end{document}